\documentclass[prl,twocolumn,a4paper,amsmath,floatfix]{revtex4-1}
\usepackage[utf8]{inputenc}
\usepackage{graphicx}
\usepackage{amsmath}
\usepackage{epstopdf}
\usepackage{url}
\usepackage{hyperref}
\usepackage{color}
\begin{document}
\title{Tuning the liquid-liquid transition by modulating the hydrogen bond angular flexibility in a model for water}
\author{Frank Smallenburg}
\affiliation{Institut f\"ur Theoretische Physik II: Weiche Materie, Heinrich-Heine Universit\"at D\"usseldorf, Universit\"atstrasse 1, 40225 D\"usseldorf, Germany}
\author{Francesco Sciortino}
\affiliation{Department of Physics and CNR-ISC, Sapienza, Universit$\acute{a}$ di Roma, Piazzale Aldo Moro 2, I-00185, Roma, Italy.}

\date{\today}

\begin{abstract}
We propose a simple extension of the well known  ST2 model  for water  [F.H. Stillinger and A. Rahman, J. Chem. Phys. {\bf 60}, 1545 (1974)] that allows for a continuous modification of the 
hydrogen bond angular flexibility.  We show that  the bond flexibility  
affects the relative thermodynamic stability of the liquid and of the hexagonal (or cubic) ice.
On increasing flexibility, the  liquid-liquid critical point, which in the original ST2 model is located in the no-man's land (i. e. the region where
ice is the thermodynamically stable phase)  progressively moves to a temperature 
 where the liquid is more stable than ice.   Our study definitively proves that the liquid-liquid transition in ST2 is a genuine phenomenon, of high relevance in all tetrahedral network-forming liquids, including water.
 \end{abstract}

\maketitle
The possibility that a one-component system assumes (beside the gas phase) more than one
disordered condensed phase is currently highly debated in liquid state physics.  Since the original proposition~\cite{poole1992phase} 
of a liquid-liquid  (LL) transition in supercooled water (based on a molecular dynamics study of the ST2~\cite{stillinger1974improved}  potential),  a large literature body has investigated this topic, suggesting that the microscopic  origin  of  a LL transition must be attributed to the competition between   two  local structures,
differing in  energy, entropy and density~\cite{debenedettinovel, debenedetti2003supercooled, mishima1998relationship, amann2013water,  
fuentevilla2006scaled,
cuthbertson2011mixturelike,
  taschin2013evidence,franzese2001generic,tanaka,gallo2012ising,russoj,holten2014two}.  Still, when and how the interaction potential
between  molecules will favor a LL transition  which can be accessed in the absence of crystallisation
is rather unclear.  Only recently an effort has been made to provide a picture which simultaneously accounts for the free energy of both
the liquid and ordered phases~\cite{limmerchandler2011, palmer2014metastable, natphys2},  as well as of the kinetic barrier separating them. 
The driving force behind this renewed interest in the physics of LL transitions has been provided by two very 
controversial studies from the same group~\cite{limmerchandler2011,limmerchandler2013}. These studies  state that
in previous numerical investigations --- including  the ST2-based results  which  originated the  LL transition concept  --- 
the low density liquid phase appearing below the LL critical point  was in reality an ice phase, i.e.
 crystallization was mistaken for a LL transition.  
Several following investigations by different groups have disagreed with this interpretation, providing further support in favour of the presence of two well-defined  distinct liquid phases in the ST2 model~\cite{sciortino2011study, poole2013free, kesselring2012nanoscale, liu2012liquid,palmer2014metastable}. 

The most recent contribution~\cite{palmer2014metastable}  has  confirmed that the   free energy basins of the two liquids are well separated from the crystal one and hence, in principle, both liquid phases can be explored in metastable equilibrium conditions. Of course, this requires that  
 the metastable liquid phases  survive for a time longer than   the equilibration time. Such times can not be calculated by thermodynamic
 information only. It is thus worth exploring the possibility of a definitive proof of the existence of a LL critical point in ST2 
 that does not require kinetic information. 
We present such a proof here by {\it continuously} tuning one of the ST2 model parameters.
We show  that it is possible to
 modulate the relative stability of the liquid and of the hexagonal (or cubic) ice I$_{h/c}$ such that 
 the melting temperature of  I$_{h/c}$ drops below the LL critical temperature.
 Under these conditions, the low-density liquid is thermodynamically more stable than I$_{h/c}$, demonstrating that  these
two phases are distinct.   The  results reported in this Letter  not only conclude once and for all the debate on the existence of a genuine LL transition in the ST2 model but also provide important clues on the mechanisms controlling crystal stability in tetrahedral lattices.

Our study builds on recent investigations of patchy colloidal particles, interacting via four attractive patches tetrahedrally  located on the particle surface\cite{romano,widepatches,natphys2}.  Searching for the particle properties favoring the self-assembly of  the technologically relevant diamond lattice~\cite{maldovan},  it has been discovered that very flexible bonds destabilize open crystal phases so much that the liquid  retains its thermodynamic stability even at very low temperatures~\cite{widepatches}. 
Crystallization is instead favored  by highly directional bonds.  In addition, at low temperature $T$ these systems can exhibit a LL transition between two interpenetrating tetrahedrally coordinated networks~\cite{natphys2}.  On increasing the bond flexibility the LL transition becomes thermodynamically stable.  These results have been confirmed in another colloidal model mimicking DNA constructs with valence four~\cite{starrsoftmatter}. Both colloidal models are characterized by a very large interparticle softness, allowing for full network interpenetration~\cite{starrinterpenetrating}.  As a result, the density of the coexisting high density liquid approximately doubles the density of the  low  density liquid phase, a factor significantly larger than what is expected for water. 
This  extreme softness casts some doubt on the applicability of these results to molecular systems and water in particular. We alleviate these doubts here, supporting once more the
hypothesis that the  liquid-liquid transition is a genuine feature of  tetrahedral network forming liquids.

\begin{figure} 
\centerline {
\includegraphics[width=2.5in]{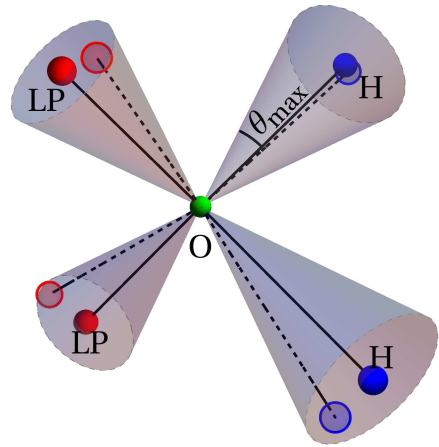}
}
\caption{Schematic plot of the ST2 water model and of the proposed extension to modulate hydrogen bond flexibility. Solid lines indicate the position of the H and LP sites in the rigid original ST2 model.  The cones have an angular amplitude equal to $\theta_\mathrm{max}$ and define the
volume limiting  the position of the same sites in the flexible model (dashed lines). 
 }
\label{fig:image}
\end{figure}

The original ST2 potential envisions a water molecule as a rigid body: the oxygen atom (O)  is located at  the centre, while the two  
protons (H) are located at a distance of 1 \AA{} from O, forming a fixed H$\hat{\rm O}$H tetrahedral angle. Two  sites (mimicking the lone-pairs, LP) are located at distance 0.8 \AA{} from O, such that the two O-H  and  the two O-LP unit vectors form the vertices of a perfect tetrahedron.
The H and LP sites carry an electric charge.  Long-range electrostatic interactions are included via the reaction-field.  
Complete details of the simulation procedure are as described in Ref.~\cite{poole1992phase}. For this model, the phase diagram has recently been evaluated, demonstrating the stability of ice I$_{h/c}$ at low temperature and pressure \cite{st2phase}, consistent with the recent observation of ice I in simulation of ST2~\cite{yagasaki2014spontaneous}.
We  modulate the flexibility of the hydrogen bonds by allowing the unit vectors pointing toward the  H and LP sites to fluctuate (with no additional energy cost) with respect to the original direction, with a maximum angle $\theta_\mathrm{max}$ (see Fig. \ref{fig:image}).  By changing $\theta_\mathrm{max}$ 
it is possible to continuously tune the bond flexibility.  When  $\theta_\mathrm{max}=0^\circ$ the modified model coincides with the original ST2 model. 
Apart from   $\theta_\mathrm{max}=0^\circ$ ($\cos\theta_\mathrm{max}=1.0$), we explore in detail the cases
$\theta_\mathrm{max}=8.11^\circ$ ($\cos\theta_\mathrm{max}=0.99$),  $\theta_\mathrm{max}=11.5^\circ$ ($\cos\theta_\mathrm{max}=0.98$)
and the case   $\theta_\mathrm{max}=14^\circ$ ($\cos\theta_\mathrm{max}=0.97$). We note that in principle, an energy cost to bending could be included in the model. However, the main effect of this would be to make the effective bond flexibility temperature-dependent, and we have thus omitted this here.

\begin{figure} 
\centerline {
\includegraphics[width=2.8in]{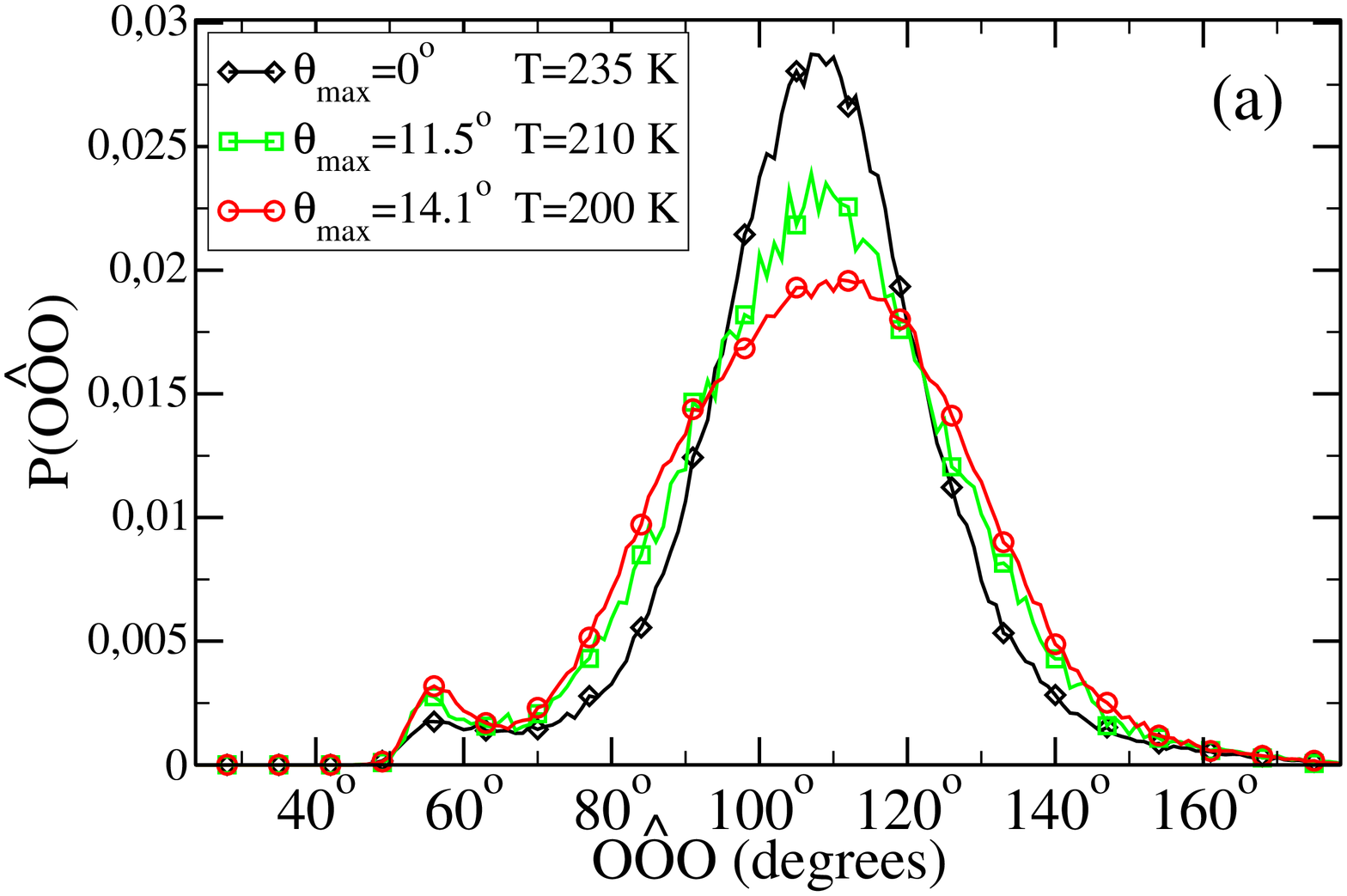}
}
\centerline {
\includegraphics[width=2.8in]{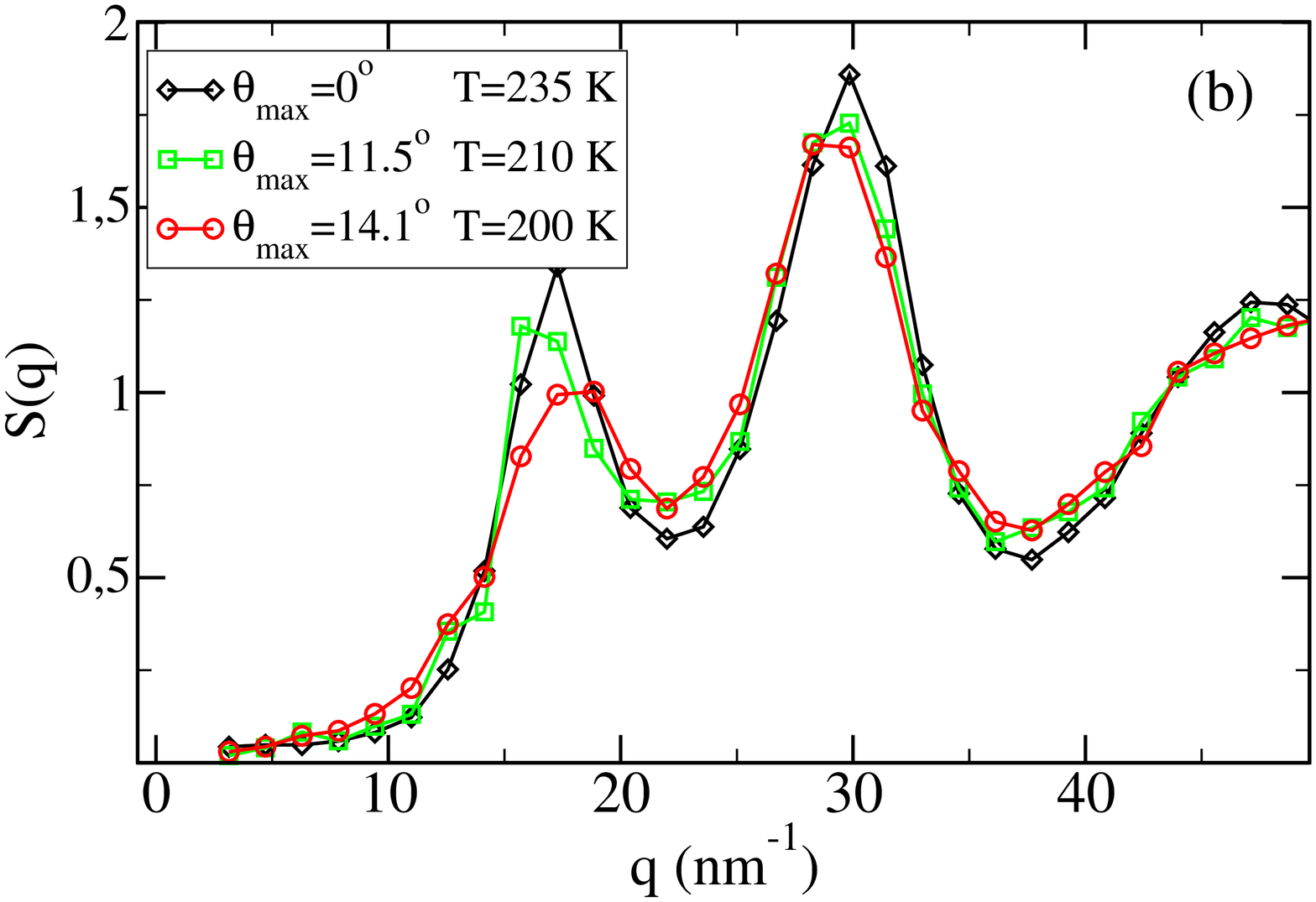}
}
\caption{(a) Probability distribution of the O$\hat{\rm O}$O angle for bonded triplets at $\rho=0.90$ g/cm$^3$ for three different values of the flexibility.  Note the increase of the variance of the distribution on increasing flexibility. Two adjacent water molecules are considered bonded if the 
OO distance is less than 3.2 \AA. (b)  Structure factors $S(q)$ for the same state points, displaying the effect of the flexibility on the pre-peak. }
\label{fig:geo}
\end{figure} 

To provide evidence that on increasing $\theta_\mathrm{max}$, the tetrahedral network becomes more and more flexible  we  evaluate the  O$\hat{\rm O}$O angle distribution $P$(O$\hat{\rm O}$O) between bonded triplets and the structure factor $S(q)$, at the lowest $T$ we have been able to
equilibrate and at the optimal network density. 
Previous studies have shown that the width of $P$(O$\hat{\rm O}$O)   as well as the 
amplitude of the pre-peak in  $S(q)$ correlate with  bond flexibility~\cite{saika2013understanding}.
Fig.~\ref{fig:geo} shows that the angular distribution is  centred around the tetrahedral angle but  widens  on increasing
$\theta_\mathrm{max}$, indicating the larger number of geometrical
arrangements available for the formation of the network.  Simultaneously,  the larger disorder in the network structure decreases the
intensity of the $S(q)$ pre-peak, in full agreement with results based on tetrahedral patchy colloids~\cite{saika2013understanding}.

\begin{figure} 
\centerline {
\includegraphics[width=2.5in]{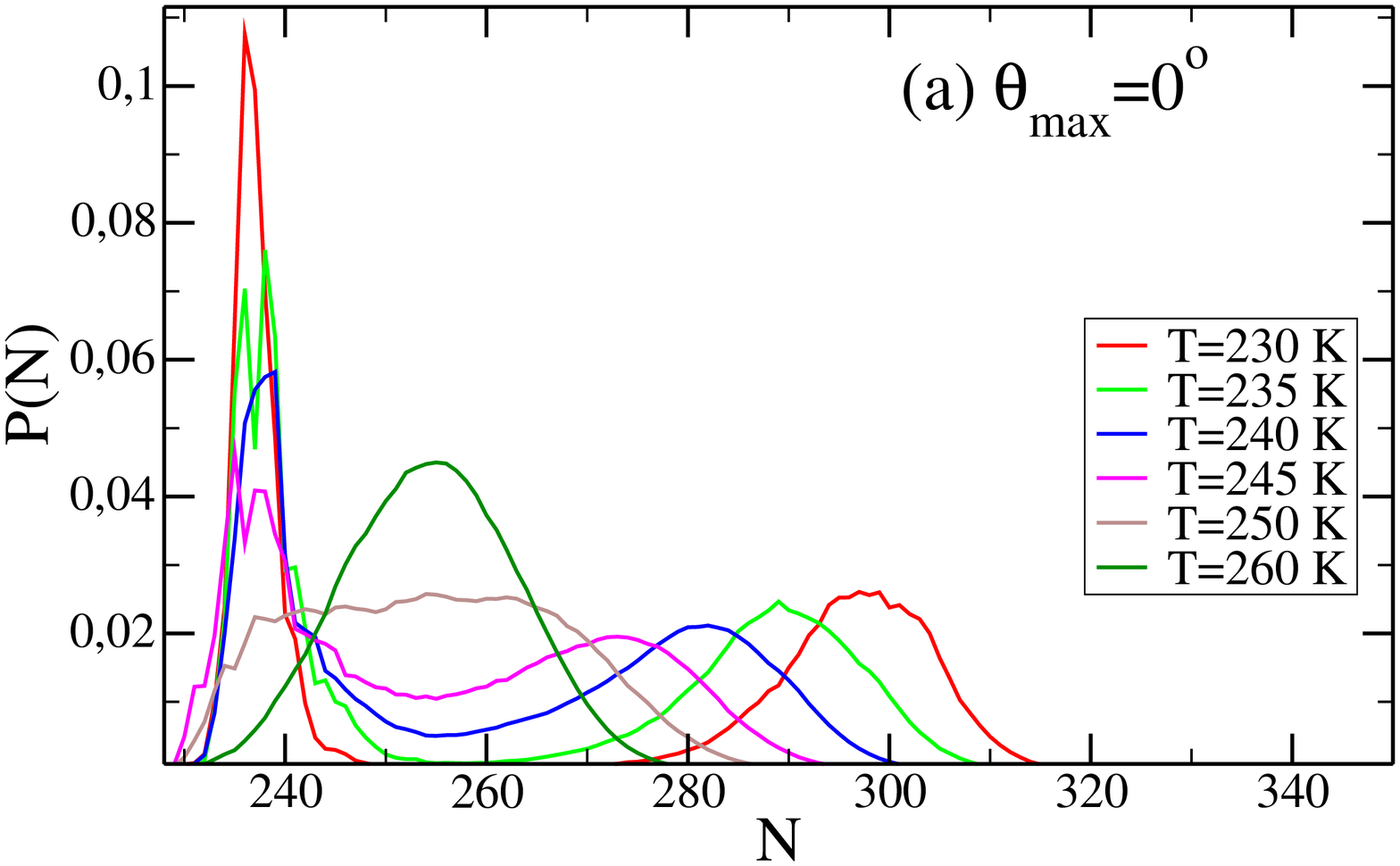}}
\centerline {
\includegraphics[width=2.5in]{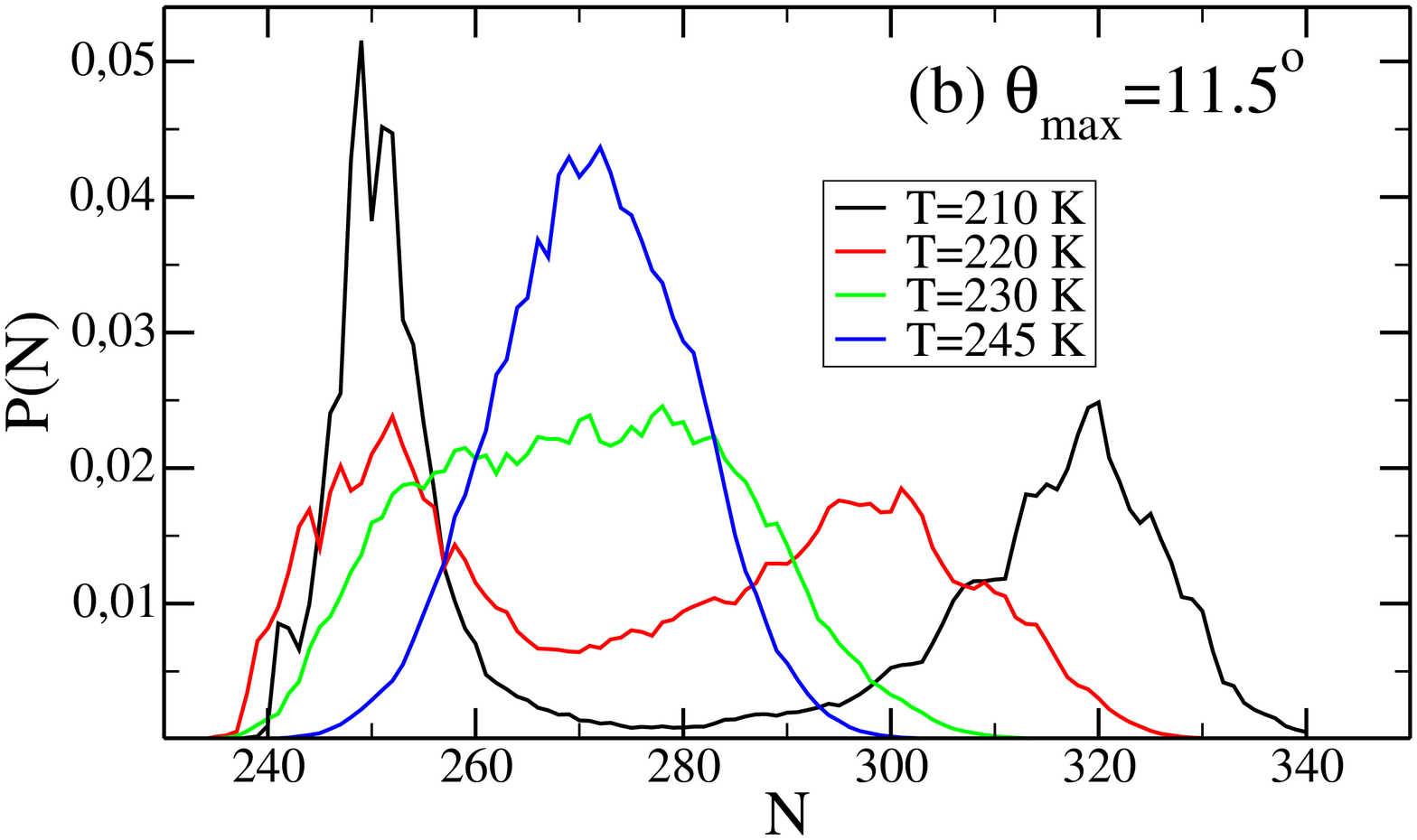}}
\centerline {
\includegraphics[width=2.5in]{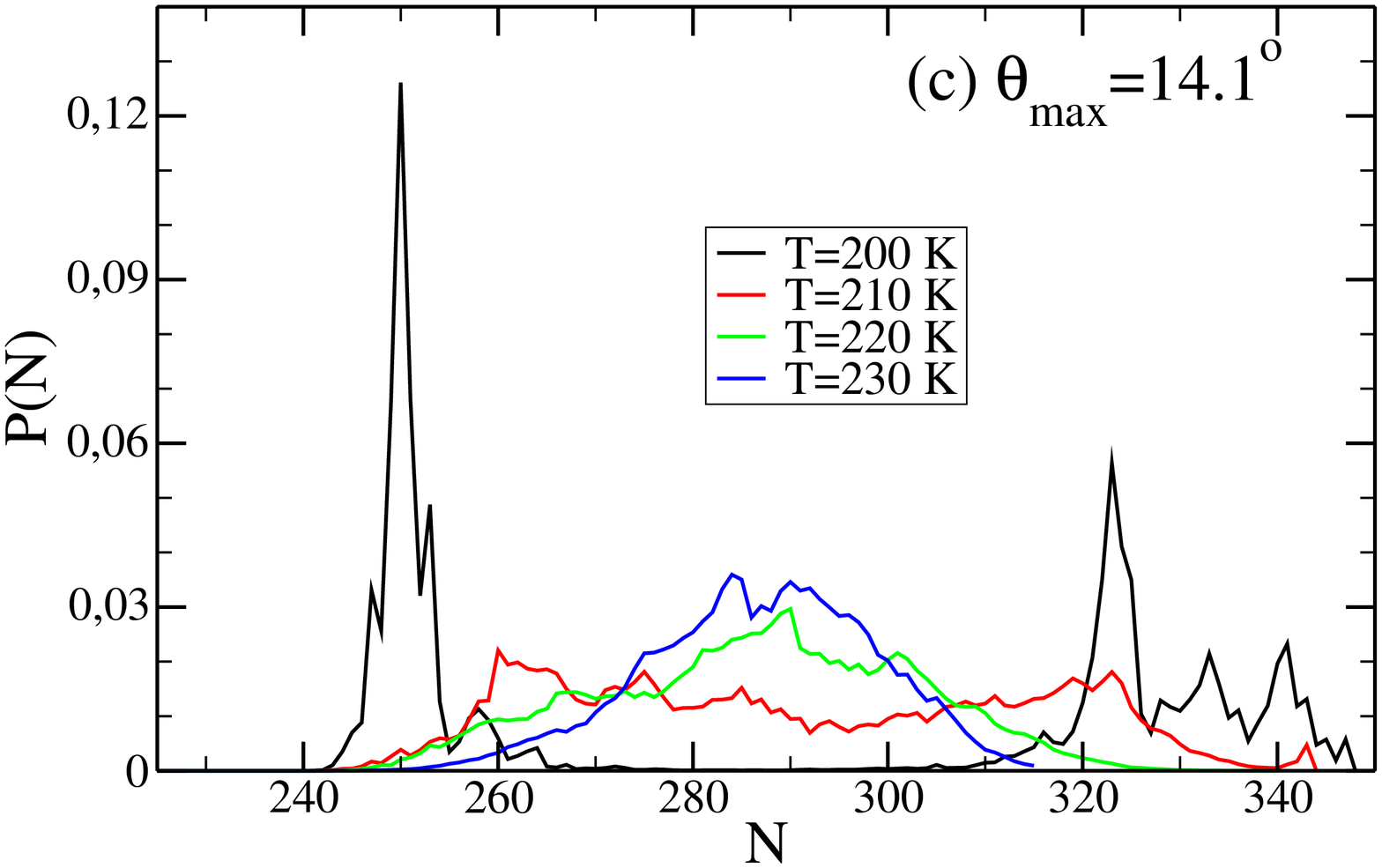}
}
\caption{Distribution  $P(N)$ in the number $N$ of particles populating a volume of 8 nm$^3$ at fixed temperature and chemical potential $\mu$.
This last quantity controls the average density and it is selected to provide equal area below the low-density and high-density liquid
phases. $P(N)$ evolves from a single-peak to a double peak shape on crossing $T_c^{LL}$. The data for $cos\theta_\mathrm{max}=1$ are redrawn from
Ref.~\cite{sciortino2011study}.
}
\label{fig:pofN}
\end{figure}

\begin{figure} 
\centerline {
\includegraphics[width=3.2in]{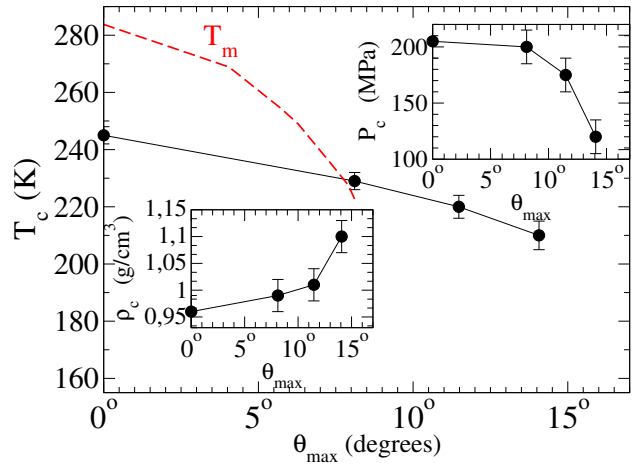}
}
\caption{Dependence of the liquid-liquid critical point temperature $T_c^{LL}$ on bond flexibility calculated from free energy estimate based
on successive umbrella sampling simulations. These grand-canonical simulations assume a volume of 8 nm$^3$.  The dashed red line
indicates  $T_m$, the melting $T$ for the liquid to ice I$_{h/c}$ transformation at the critical pressure.  The two insets show respectively the critical pressure $P_c$ and the critical density $\rho_c$ as a function of   $\theta_\mathrm{max}$. 
} 
\label{fig:tcvscosteta}
\end{figure}

To estimate the location of the LL critical point we  perform  grand-canonical Monte Carlo simulations for different $T$ to 
estimate the probability $P(N)$ of finding $N$ particles in the simulated volume $V$ (8 nm$^3$).  By implementing the successive umbrella sampling technique~\cite{virnau2004calculation}, we have distributed the evaluation of $P(N)$ over multiple processors,  each of them evaluating the ratio
$P(N+1)/P(N)$,  for $220 < N < 350$.  More than 1000 processors running full time have been dedicated for six months to these calculations. 
Close to a second-order critical point, $P(N)$ develops a double peak structure that  becomes more and more pronounced on cooling, signalling the coexistence of phases with different density.  During all runs, we have constantly checked that the number of crystalline particles  (evaluated with the standard algorithms for detecting ice local structures~\cite{romano,bolhuis}) never
exceeded ten nor showed any trend toward growing. The results of these calculations for different $\theta_\mathrm{max}$ and $T$ are reported
in Fig.~\ref{fig:pofN}, spanning the $T$ interval over which $P(N)$ crosses from a single to a double-peaked function with 
a peak-to-valley ratio around 0.5 (the characteristic value assumed by $P(N)$ at the critical temperature~\cite{wilding1995critical}) down to $T$ where
the two peaks are well resolved, signalling the onset of a clear free energy barrier between the low-density and high-density
liquids.  The estimated location of the critical temperature $T_c^{LL}$  for the investigated box side ($L=2$ nm)  as a function of $\theta_\mathrm{max}$ 
are shown in Fig.~\ref{fig:tcvscosteta}. Consistent with what was previously found for the patchy and DNA colloidal models, increasing  bond flexibility (i.e. increasing $\theta_\mathrm{max}$) results in lowering  $T_c^{LL}$. 
Additionally, upon increasing the flexibility, the critical pressure decreases and the critical density increases, consistent with the coupling between 
bond flexibility and local density. Indeed, for tetrahedral patchy particles it has been shown that the density at zero pressure of the fully bonded network decreases with increasing bond directionality. Similarly, increasing flexibility results in a network that is much more easily compressible~\cite{saika2013understanding}. Our results suggest
that  the coupling between local density, compressibility and flexibility also affects the critical parameters.

To properly frame the LL transition in terms of thermodynamic stability compared to  I$_{h/c}$ we evaluate the free energy of the liquid and of the two ices.   To evaluate the liquid free energy, we perform thermodynamic integration from the ideal gas~\cite{st2phase}. To evaluate the crystal free energy
we integrate from an Einstein crystal in the molecular framework~\cite{vega2008determination}, extending the method to account for
the flexible arms.   For this, we use as a reference system a thermalised  ice I$_h$ or I$_c$ configuration with at least 20000
particles to average over proton-disorder.  For each molecule in the reference system we define  the
reference oxygen position (${\bf r}_0$), the reference orientation of the dipole and HH unit vectors and the
reference orientation of the OH and OLP unit vectors, all in  the ideal tetrahedral geometry. 
For each particle, the reference Hamiltonian consists of three parts:
\begin{eqnarray}
H_\mathrm{trans}  &=& \lambda_t (\left|\mathbf{r} - \mathbf{r}_0\right|)^2/\sigma^2\\
H_\mathrm{rot}  &=&\lambda_r \left[\sin^2\phi_a + \left(\frac{\phi_b}{\pi}\right)^2 \right],\\
H_\mathrm{arms}  &=& \lambda_a \sum_{i=1}^4 [1 - \cos \theta_i]  
\end{eqnarray}
where $\left|\mathbf{r} - \mathbf{r}_0\right|$ is the distance between the position of the oxygen atom in the reference and in the instantaneous configuration, $\sigma=1 $nm is a convenient length scale, 
$\phi_a$ and $\phi_b$ are respectively the angle between the reference and the instantaneous position of the ideal dipole and HH unit vectors and $\theta_i$ is the angle between the instantaneous position of the $i$ patch unit vector and the ideal position of that unit vector. In other words, this is the angle we limit in our model when specifying $\theta_\mathrm{max}$.

Monte Carlo moves, preserving the centre of mass position~\cite{frenkel2001understanding}  are performed by randomly translating a molecule,  
rotating a single patch (which changes only $H_\mathrm{arms}$), or rigidly rotating the water molecule (which changes only $H_\mathrm{rot}$).

The reference free energy (per particle) of the fully constrained system (limit of high $\lambda$) is (with $\beta=1/k_BT$ and $k_B$ the Boltzmann constant):
\begin{equation}
\beta f_\mathrm{ref} = \beta f_\mathrm{trans} + \beta f_\mathrm{rot} + 4 \beta f_\mathrm{arm},
\end{equation}
with
\begin{eqnarray}
 \beta f_\mathrm{trans} &=& -\frac{1}{N} \log\left[\left(\frac{\pi}{\beta \lambda_t}\right)^{3(N-1)/2} N^{3/2} \frac{V}{\sigma^3 N}\right],\\
 \beta f_\mathrm{rot}   &=& -\log\left[\sqrt{\pi}/4\right] + \frac{3}{2} \log (\beta \lambda_r ) \\
 \beta f_\mathrm{arm}   &=& 
                  -\log \left[\frac{(1-\exp[-\beta \lambda_a(1-\cos\theta_\mathrm{max})])}
                  {(1-\cos \theta_\mathrm{max}) \beta \lambda_a }  \right] \\
                  &\simeq& \log ((1-\cos \theta_\mathrm{max}) \beta \lambda_a)
\end{eqnarray}

\begin{figure} 
\centerline {
\includegraphics[width=3.0in]{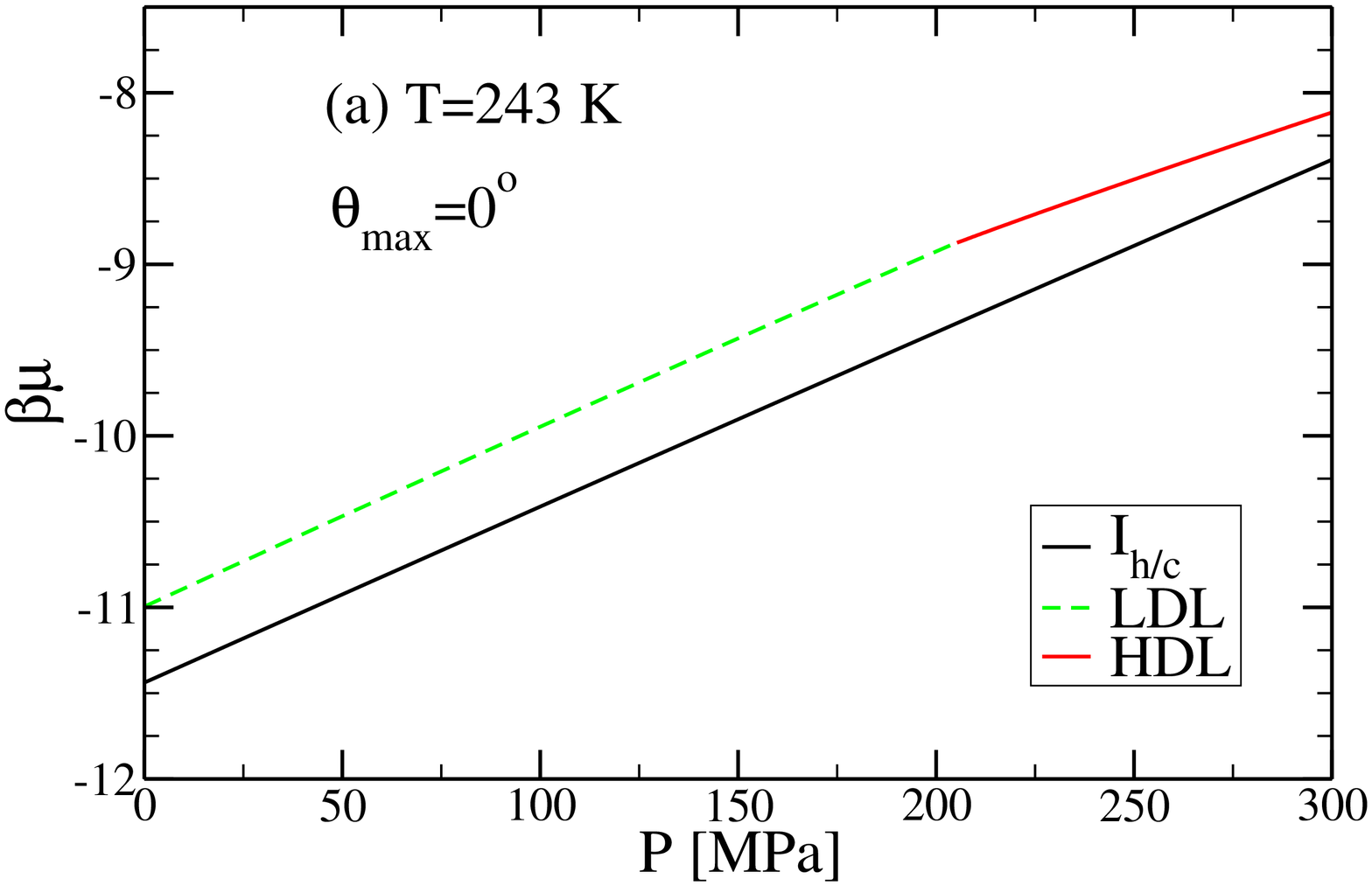}}
\centerline {
\includegraphics[width=3.0in]{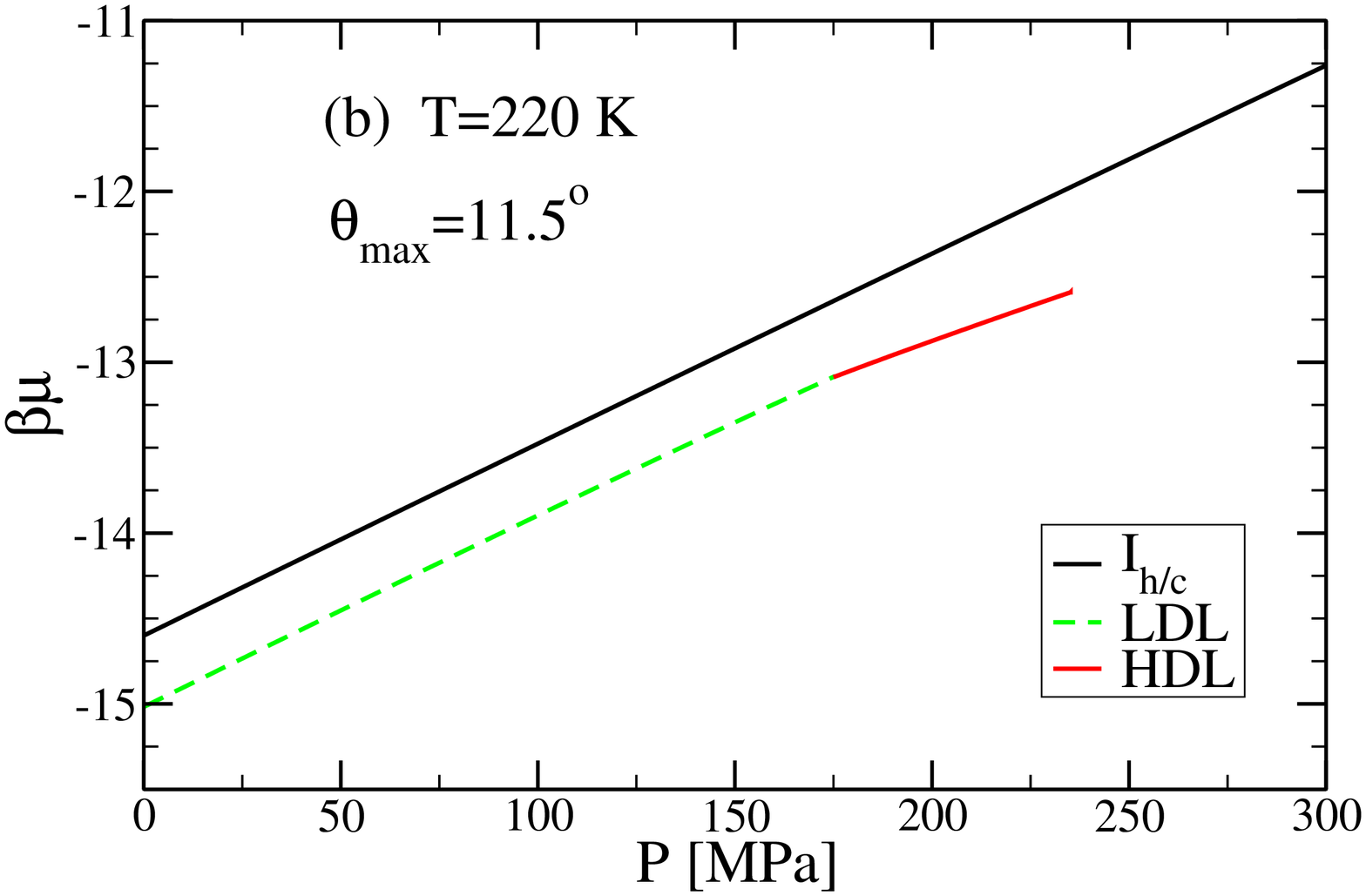}}
\caption{Pressure $P$ dependence of the reduced chemical potential $\beta \mu$ for ice I$_{h/c}$ and for the low and high
density liquid phases at $T_c^{LL}$ for (a) $\theta_\mathrm{max}=0^\circ$ and (b) $\theta_\mathrm{max}=11.5^\circ$.  Note that while for $\theta_\mathrm{max}=0^\circ$ (e.g. for the original ST2 model)  $\beta \mu$ of the liquid phases
is higher than the one of ice  I$_{h/c}$, the opposite is found already for  $\theta_\mathrm{max}=11.5^\circ$.  Under these conditions, the liquid is thermodynamically more stable than I$_{h/c}$. }
\label{fig:bmu}
\end{figure}

The model free energy $f$ is then calculated following the methodology described in Ref.~\cite{vega2008determination}.
As for the original ST2 model,  at all temperatures and for all  values of $\theta_\mathrm{max}$, 
we find that  ice I$_h$ and I$_c$  have the same free energies, within our numerical precision (with an uncertainty in $\beta f$ of $ \pm 0.01)$. 
From the free energy and the equation of state we evaluate the chemical potential $\beta \mu= \beta f + \beta P/\rho$, where
$P$ is the pressure and $\rho$ the number density. The main results of the Letter are reported in Fig.~\ref{fig:bmu}, showing the $P$ dependence of the liquid and ice I$_{h/c}$  chemical potential at the LL critical temperature $T_c^{LL}$.  In the case of the original
ST2 model ($\theta_\mathrm{max}=0^\circ$), $\beta \mu$  of  I$_{h/c}$  is always lower than the liquid one,
consistent with the location of $T_c^{LL}$ in the no-mans land.     On increasing  $\theta_\mathrm{max}$, the relative
stability  of the  I$_{h/c}$  compared to the liquid  changes. At  $\theta_\mathrm{max} \simeq 8^\circ$, the
liquid chemical potential is slightly  lower than  ice I$_{h/c}$ , while for $\theta_\mathrm{max}=11.5^\circ$, at $T_c^{LL}$
the liquid phase has gained a significant stability compared to the open crystal lattice. Since the crystal free energy
is higher than the liquid one (already for $\theta_\mathrm{max}\simeq 8^\circ$ !)  there is no possibility that the low density liquid phase in  flexible ST2 will ever convert into  the  I$_{h/c}$ structure.  The low-density liquid phase is, without any ambiguity, a phase by itself, definitively disproving the
arguments in Ref.~\cite{limmerchandler2011,limmerchandler2013}. 
In an expanded representation of the phase diagram, in which we include  $T$, $P$ and $\theta_\mathrm{max}$, the lines of LL critical points  moves with continuity  from a condition of metastability with respect to $I_h$ and $I_c$ to a condition of stability, around $\theta_\mathrm{max}=8^\circ$. This continuity allows us to conclude that the LL critical point observed in the original ST2 model must also be genuine. We stress that our study does not aim at providing a (better or worse) model for water but to show --- with an extremely simple modification to the ST2 model --- that the liquid-liquid transition can be made thermodynamically stable. 
For the case of water, the estimated LL critical point is definitively located in the region in which ice nucleation in bulk is dominant.  There, only ingenious  experiments in the negative pressure region of the phase diagram~\cite{azouzi2013coherent,pallares2014anomalies} or in the low $T$ glass phases~\cite{amann2013water} can  provide important clues. 
Still, our proof reinforces the idea that the LL transition is a genuine phenomenon in all 
tetravalent systems for which a suitable softness allows for network interpenetration and a suitable bond flexibility allows for
enhanced  stability of the liquid phase(s) compared to open crystal structures.

%

 \end{document}